\documentclass[lettersize,journal]{IEEEtran}
\usepackage{amsmath,amsfonts}
\usepackage{algorithmic}
\usepackage{algorithm}
\usepackage{array}
\usepackage[caption=false,font=normalsize,labelfont=sf,textfont=sf]{subfig}
\usepackage{textcomp}
\usepackage{stfloats}
\usepackage{url}
\usepackage{verbatim}
\usepackage{graphicx}
\usepackage{cite}
\usepackage{xcolor}
\DeclareMathOperator{\diag}{diag}
\def\vect{\mathbf}
\def\matr{\mathbf}
\newcommand\norm[1]{\left\lVert#1\right\rVert}

\begin{document}
\bstctlcite{IEEEexample:BSTcontrol}
    \title{Deep Mismatch Channel Estimation in IRS based 6G Communication
\thanks{Swapnil~Saha and       Md.~Forkan~Uddin are with the Department of Electrical and Electronic Engineering, Bangladesh University of Engineering and Technology.}

    }
  \author{Swapnil~Saha and Md.~Forkan~Uddin}




\maketitle

\begin{abstract}

We propose a channel estimation protocol to determine the uplink channel state information (CSI) at the base station for an intelligent reflecting surface (IRS) based wireless communication. More specifically, we develop a channel estimation scheme in a multi-user system with high estimation accuracy and low computational complexity. One of the state-of-the-art approaches to channel estimation is the deep learning-based approach. However, the data-driven model often experiences high computational complexity and, thus, is slow to channel estimation. Inspired by the success of utilizing domain knowledge to build effective data-driven models, the proposed scheme uses the high channel correlation property to train a shallow deep learning model. More specifically, utilizing the one coherent channel estimation, the model predicts the subsequent channel coherence CSI. We evaluate the performance of the proposed scheme in terms of normalized mean square error (NMSE) and spectral efficiency (SE) via simulation. The proposed scheme can estimate the  CSI with reasonable success of lower NMSE, higher SE, and lower estimation time than existing schemes.  
\end{abstract}


%
\IEEEpeerreviewmaketitle


\section{Introduction}

\IEEEPARstart{R}{ecently}, intelligent reflecting surface (IRS), which has the capability of shaping the wireless
channels between the users and the base station (BS) to enhance the system performance, has been proposed as a promising technology for the future wireless communications, e.g., $6$G communication \cite{bariah2020prospective,long2021promising}. 
For high-frequency communication, like $6$G, where the attenuation loss is high and the line of sight (LOS) component is weak, IRS can create a virtual LOS path by reflecting the incident signal from the BS with a proper phase difference \cite{tan2018enabling}. It has been considered a promising technology to overcome the spectrum and energy efficiency challenges of the $6$G network \cite{wu2021intelligent,wu2019towards,naeem2022irs}. As it has gained popularity, extensive work has been done on IRS optimization for a given perfect channel state information (CSI). However, in practical implementation, before optimizing the IRS and data transmission, one needs to estimate the CSI between the BS and the users. Most existing optimization works are based on the assumption of perfect CSI, which is inappropriate in the practical sense \cite{zheng2022survey}. Nevertheless, gaining the exact CSI is not easy; it has to consider different factors, like the noise of the environment \cite{liu2021deep}, low estimation time for a practical number of users, and the number of IRS elements \cite{zheng2020fast}. 
Motivated by the above practical issues, we aim to develop a channel estimation scheme that can provide CSI with high accuracy, especially in noisy environments, utilizing deep learning supremacy with minimum training parameters.\\
There is a substantial amount of literature on deep learning-based channel estimation in IRS-based wireless systems. (\cite{liu2021deep,elbir2020deep,liu2020deep,liu2020deepofdm,liu2022deepISAC} and references therein). In line with our work on multi-user communication (IRS-MUC) systems, the basic approach is to formulate the channel estimation problem as a de-noising problem and train the data-driven model with high accuracy. For instance, \cite{liu2021deep} proposed a DL-based channel denoising algorithm that improves the noisy least square (LS) channel estimator. N. K. Kundu et al. expressed the channel estimation problem as an image-denoising problem and solved it by designing a twin convolutional neural network (CNN) \cite{elbir2020deep}. This work\cite{liu2020deep} proposed a similar problem formulation with a hybrid IRS configuration to reduce the training overhead.\\
However, there is a trade-off between the computational complexity (thus inference time of CSI estimation) and performance; the data-driven model performs better compared to the traditional approaches at the expense of high computational complexity \cite{liu2021deep}. On the contrary, in the high transmission data protocol, it is desired that minimum time will be spent estimating CSI. Motivated by this practical importance, we aim to estimate CSI by a data-driven model with a possible minimum number of parameters to reduce the computational complexity. The reward for the lower computational complexity comes from using the high correlation channel character to build the data-driven model. To understand the scenario, consider the context where CSI changes slowly from one coherence time to another coherence time. Because of the existing correlation between them, it is possible to estimate subsequent CSI if you have the prior CSI estimation and knowledge of the necessary adjustment/mismatch. We aim to learn that mismatch through our shallow, deep-learning model. Our approach is also inspired by works that utilize domain knowledge to enhance the performance of data-driven models. \cite{regenwetter2022deep,ozbayoglu2020deep,liu2020k,rajpurkar2017chexnet,muralidhar2018incorporating}.


The rest of the letter is organized as follows. Section \ref{System Model} describes the system model and problem formulation. The proposed method for finding the channel mismatch and channel estimation is described in Section \ref{Problem Formulation}. The simulation results are presented in Section \ref{simulation}. Section \ref{Conclusion} concludes the letter. 


\begin{figure*}[!t]
\centering
\subfloat[]{\includegraphics[scale=0.15]{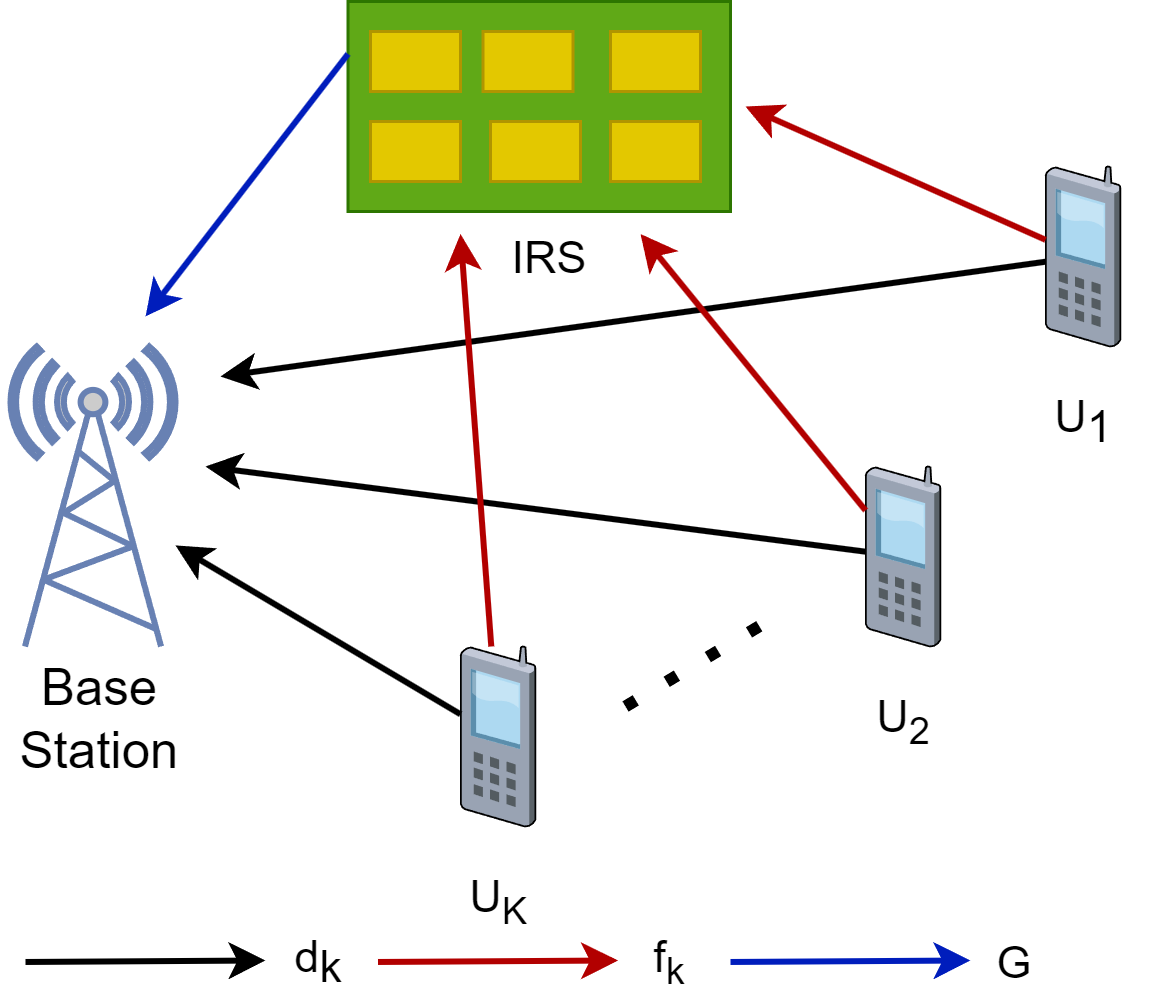}%
\label{fig:IRS_MUC_Uplink}}
\hfil
\subfloat[]{\includegraphics[scale=0.15]{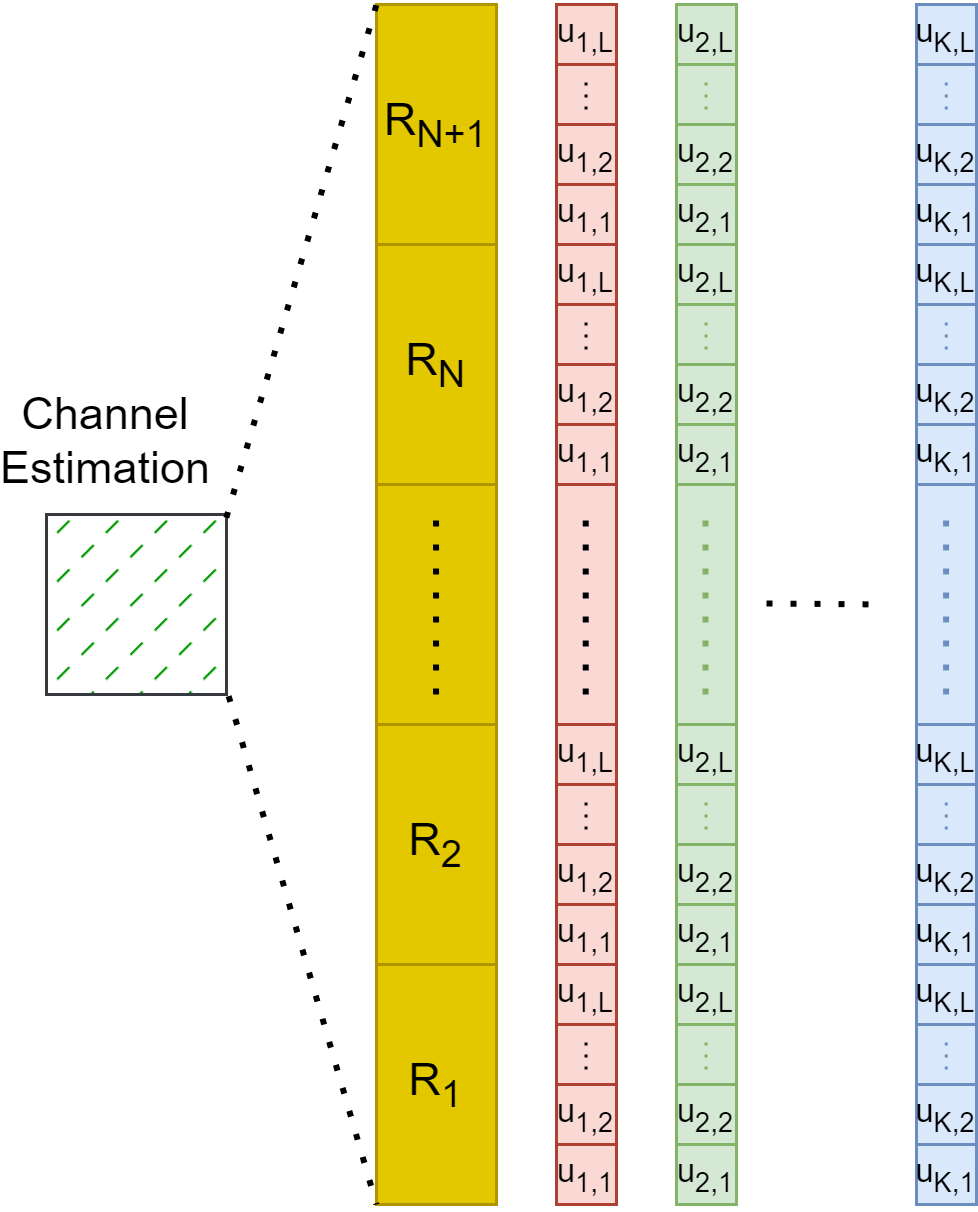}%
\label{fig_second_case}}
\caption{(a) System architecture for uplink channel estimation, (b) Pilot signal transmission in channel estimation phase.}
\label{fig:Pilot Signal}
\end{figure*}

\section{System Model and Problem Description}\label{System Model}
We consider a single IRS for a communication system with $K$ users and one base station (Multiple Input Single Output (MISO)), as shown in Fig. \ref{fig:IRS_MUC_Uplink}. The set of the users is denoted by $\mathcal{K}=\{U_1, U_2, \ldots,U_K\}$. The base station (BS) consists of $M$ antennas, while each user has a single antenna. The IRS consists of $N$ reflecting elements. 


Let the reflection column vector be $\vect{r}= [\mathcal{B}_1 e^{j \theta_1}, \mathcal{B}_2 e^{j \theta_2}....\mathcal{B}_N e^{j \theta_N}]^T$, where $\mathcal{B}_n$ is the amplitude and $\theta_n$ is the phase shift of the $n$-th element in the IRS. The diagonal reflection matrix $\matr{R} = \diag({\vect{r}}) \in \mathcal{C}^{ N \times N}$ stores the phase state of the reflection surface. Let denote the channel gain from user $U_k$ to IRS by $\vect{f_k}$. As $\diag(\vect{r})\vect{f_k} = \diag(\vect{f_k})\vect{r} $, the  channel gain from user $U_k$ to BS through IRS can be written as $\matr{G} \diag({\vect{r}}) \vect{f_k} = \matr{G} \diag({\vect{f_k}}) \vect{r} $, where $\matr{G}$ is the channel gain from IRS to BS. The uplink CSI for a user $U_k$ can be written as $\matr{H_k} = [\vect{d_k}, \matr{B_k}] \in \mathcal{C}^{M \times (N+1)}$, where  $\vect{d_k}$ is the channel gain of the direct path between user $U_k$ and the BS, and
$\matr{B_k} = \matr{G} \diag({\vect{f_k}}) \in \mathcal{C}^{M \times N}$ is the cascaded channel gain from user $U_k$ to BS through IRS. We consider a time-division duplex system for channel estimation, Where a time period $T$ is divided into channel estimation and data transmission. This study aims to estimate CSI, which we donate as $\hat{\matr {H}}_k$, with high accuracy and low estimation latency. Although we provide a scheme to estimate the uplink CSI, the downlink CSI can be obtained by the conjugate transpose of the uplink channels via exploiting the channel reciprocity as the time division duplex (TDD) system is assumed \cite{zheng2022survey, 9505267}.

\subsection{Channel Estimation Protocol}

The uplink channel estimation protocol is described in Fig. \ref{fig:Pilot Signal}(b). In more detail, $K$ different orthogonal pilot signals, each with length $L (L > K)$, are transmitted from the $K$ users. The pilot signal of a user $U_k \in \mathcal{K}$ is given by $\vect{u}_k = [u_{k,1},u_{k,2},...u_{k,L}]$ where, $\vect{u_i}^H \vect{u_j} = P_tL  \quad \text{if} \quad i=j$ and $0$ otherwise.
Here, the power of each symbol is $P_t$. The channel estimation steps consist of $C$ sub-frames with $C=N+1$. In each sub-frame, the different reflection patterns are set in $C$ sub-frames that are presented as $ \matr{\wedge} = [\matr{R_1}, \matr{R_2},....\matr{R_C}]^T$. The reflection phases are kept constant in each sub-frame and changed in the next sub-frame. Each user $U_k \in \mathcal{K}$ sends their same pilot signal $\matr{u_k}$ in each subframe. Thus, the $l$-th, $l \in \{1,2,3,\ldots,L\}$ received pilot signal at the BS in the $c$-th sub frame, $\vect{s_{c,l}} \in \mathcal{C}^{M \times 1}$, can be written as $\vect{s_{c,l}}=\sum_{k=1}^K \matr{H}_k \vect{p_c} u_{k,l} + \vect{n}_{c,l}$, where, $\vect{n}_{c,l} \in \mathcal{C}^{ M \times 1}$ is complex white Gaussian noise and $\vect{p_c}=[1, \vect{r_c}]^T  \in  \mathcal{C}^{(N+1) \times 1}$. The selection of choice $N$, $K$, and $L$ depends on the system size and available energy at the BS. Nevertheless, our proposed algorithm is independent of the choice of those parameters.  

%


Now combining $L$ length pilot signals, we can write the received signal at BS at $c$-th subframe, $\matr{S_c} \in \mathcal{C}^{ M \times L}$, as follows:
\begin{equation}\label{received_pilot}
    \matr{S_c} =  \sum_{k=1}^K \matr{H_k} \vect{p_c} \matr{u_k}^T + \matr{N_c}.
\end{equation}


\subsection{Problem Formulation}
Multiplying both sides in \eqref{received_pilot} by $\vect{u_k}^*$ and using the orthogonal relationship of the pilot signal, we obtain,
\begin{equation} \label{received_pilot_matrix}
    \begin{split}
        \matr{S_c} \vect{u_k}^* & =  \sum_{k'=1}^K \matr{H_{k'}} \vect{p_c} \matr{u_k}^T \vect{u_k}^*+ \matr{N_c} \vect{u_k}^* \\
        & = P_tL \matr{H_k} \vect{p_c} + \matr{N_c} \vect{u_k}^*.
    \end{split}
\end{equation}
Let $\vect{x_{c,k}} = \frac{1}{P_t L} \matr{S_c} \vect{u_k}^* \in \mathcal{C}^{M \times 1}$ and $\vect{z_{c,k}} = \frac{1}{P_t L}  \matr{N_c} \vect{u_k}^* \in \mathcal{C}^{M \times 1}$. 
Thus, \eqref{received_pilot_matrix} can be written as $\vect{x_{c,k}} = \matr{H_k} \vect{p_c} + \vect{z_{c,k}}$.
Finally, after $C$ sub frames, we can write the rece ved signal, $\matr{X_k} \in \mathcal{C}^{M \times C}$ in the following matrix format
\begin{equation} \label{target euation}
    \matr{X_k} = \matr {H_k} \matr{P} + \matr{Z_k},
\end{equation}
where, $\matr{P} = [\vect{p_1}, \vect{p_2}...\vect{p_C}] \in \mathcal{C}^{(N+1) \times C}$ with $\vect{p_c} = [1, \vect{r_C}]$ and $\matr{Z_k} = [\vect{z_{1,k}},\vect{z_{2,k}}..\vect{z_{C,k}}] \in \mathcal{C}^{M \times C}$.  According to \cite{jensen2020optimal}, the optimum design of $\matr{P} \in \mathcal{C}^{}$, in terms of the high received signal power at the BS, is discrete Fourier transform (DFT) matrix. As it is not possible to recover $\matr {H}_k$ directly from \eqref{target euation}, we want to estimate the CSI, denoted as $\hat{\matr {H}}_k$, with high accuracy.

\section{Proposed Channel Estimation Method}
\label{Problem Formulation}


\begin{figure}
 \centering
  \includegraphics[scale=0.10]{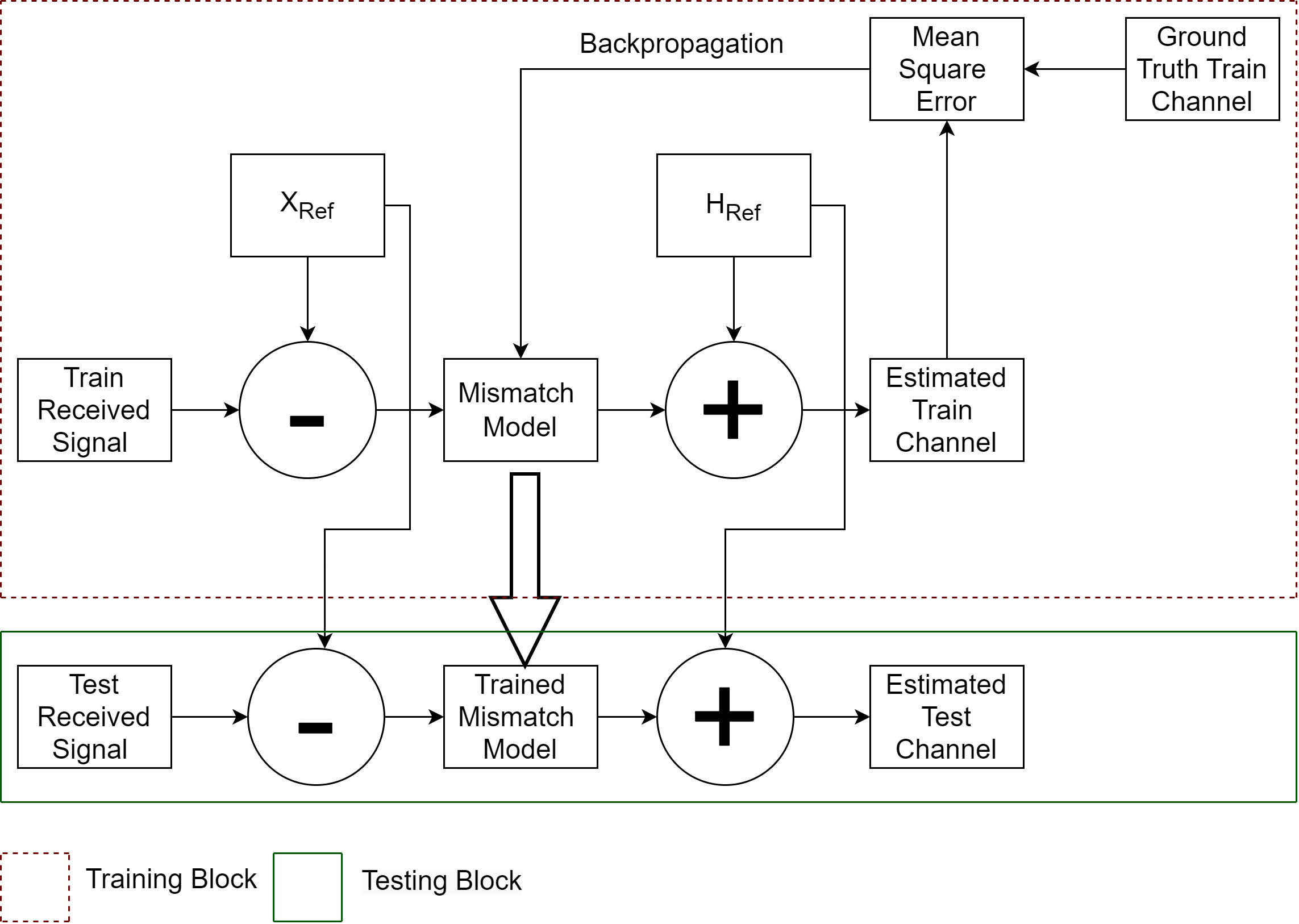}
  \caption{The block diagram of the proposed channel estimation process.}
  \label{fig:Proposed Model}
\end{figure}

 \begin{algorithm}
 \caption{Mismatch Based Channel Estimation}\label{alg:cap}
 \begin{algorithmic}[1]
 \renewcommand{\algorithmicrequire}{\textbf{Input:}}
 \renewcommand{\algorithmicensure}{\textbf{Output:}}
 \REQUIRE $\matr{X_{Ref}}$, $\matr{H_{Ref}}$, training set ($\matr{X_{Train}}$,$\matr{H_{Train}}$), convergence threshold $\eta_{threshold}$
 \ENSURE  Mismatch Model $f_{MM}()$
 \\ \textit{\textbf{Offline Training}} :
  \STATE Prepare the mismatch observation matrices $\overline{\matr{X_{Train}}}$=$\matr{X_{Train}}$-$\matr{X_{Ref}}$
  \STATE Input ($\overline{\matr{X_{Train}}}$,$\matr{H_{Train}}$) to the mismatch model $f_{MM}()$
  \STATE Estimate Mismatch $\overline{\hat{\matr{H}}_{Train}} = f_{MM}(\overline{\matr{X_{Train}}})$
  \STATE Estimate channel $\hat{\matr{H}}_{Train} = \overline{\hat{\matr{H}}_{Train}} + \matr{H_{Ref}}$
  \STATE Calculate $f_{NMSE}(\matr{H}_{Train},\hat{\matr{H}}_{Train})$
 \\ \textit{LOOP Process}
  \WHILE {Loss function does not converge} 
  \STATE Updates the parameter of the Model $f_{MM}()$
  \ENDWHILE
  \RETURN Well-trained Mismatch Model $f_{MM}()$
  \\ \textit{\textbf{Online Testing}} :
  \REQUIRE Test Sample $\matr{X_{Test}}$, trained mismatch model $f_{MM}$
 \ENSURE Estimated Channel $\hat{\matr{H}}_{Test}$
 \STATE $\overline{\matr{X_{Test}}}$ = $\matr{X_{Test}}$ - $\matr{X_{Ref}}$
 \STATE $\overline{\hat{\matr{H_{Test}}}} = f(\overline{\matr{X_{Test}}})$
 \STATE $\hat{\matr{H}}_{Test}$ = $\overline{\hat{\matr{H_{Test}}}}$ + $\matr{H_{Ref}}$
 \RETURN $\hat{\matr{H}}_{Test}$
 \end{algorithmic}
 \label{alg:algo1}
 \end{algorithm}
The proposed channel estimation is based on the fact that a sufficient correlation exists between the subsequent coherence time slots when the CSI changes slowly. The dataset used in this study (described in section \ref{simulation}) showed an average $40\%$ correlation (Pearson's Linear Correlation) between the successive coherence intervals. We aim to use this inherent coherent property of the CSI to build our data-driven model. A reasonable accurate channel estimation (refereed as a reference signal in our study) in one coherence time can be useful to estimate the next channel coherence time if a proper \emph{mismatch} is found with respect to the prior accurate channel estimation. To understand it better, consider two highly correlated signals $\matr{H_{Ref}}$ and $\matr{H_{1}}$. We can estimate the $\matr{H_{1}}$ if we know $\matr{H_{Ref}}$ by solving the linear regression problem: $\matr{H_{1}}=a\matr{H_{Ref}}+b$ by properly modeling slope $a$ and intercept $b$. Assuming the existence of a strong correlation, we allow slope parameter $a=1$ and let the model learn the intercept $b$, which we refer to as a mismatch channel vector. More specifically, we propose a data-driven model that estimates the mismatch channel vector corresponding to a prior channel estimation, which we refer to here as the reference CSI $\matr{H_{Ref}}$. The details of the selection reference CSI $\matr{H_{Ref}}$ are described in Sec.\ref{effect_ref_signal}. In the next section, we will describe the algorithmic steps for the channel estimation.
\subsection {Basic Concept}

Our goal is to train our data-driven model to learn the mapping between the mismatch of the received signal $\matr{X}-\matr{X_{Ref}}$ with the mismatch of the CSI, $\matr{H}-\matr{H_{Ref}}$. Here, $\matr{X}$ and $\matr{H}$ are arbitrary noisy received signal and CSI, respectively. $\matr{X_{Ref}}$ is the noisy reference received signal of the corresponding reference CSI $\matr{H_{Ref}}$.
To train the model, we will use the received training signal $\matr{X_{Train}}$ and ground truth CSI $\matr{H_{Train}}$ (the details of the dataset are given in Sec.\ref{simulation}). More specifically, we will train the deep learning model with ($\overline{\matr{X_{Train}}}$,$\matr{H_{Train}}$), where $\overline{\matr{X_{Train}}} = \matr{X_{Train}} - \matr{X_{Ref}}$ is the mismatch received signal. With a proper selection of training hyper-parameters and adequate model convergence, the model will learn, with reasonable accuracy, how the $\matr{H_{Ref}}$ needs to be changed (addition operation) to predict the training CSI, $\hat{\matr{H}}_{Train} = \matr{H_{Ref}} + \overline{\hat{\matr{H}}_{Train}}$. Here $\overline{\hat{\matr{H}}_{Train}}=f_{MM}(\overline{\matr{X_{Train}}})$ is the predicted mismatch CSI and $f_{MM}()$ is the well-trained mismatch model. The  loss function, $f_{NMSE}$, to train the neural network is given by
\begin{equation}\label{nmse}
    f_{NMSE}(\matr{H}_{Train}, \hat{\matr{H}}_{Train}) = \frac{\norm{\hat{\matr{H}}_{Train} - \matr{H}_{Train}}_F^2}{\norm{\matr{H}_{Train}}_F^2}.
\end{equation}
 Equ. \eqref{nmse} will also be used as the evaluation metric for the comparison with other models (Section \ref{simulation}). As for the training scheme, the same iterative procedure is executed until the NMSE converges sufficiently.\\

\subsection{Test Phase and the Algorithm}

During the testing phase, the model first takes input test mismatched received information $\overline{\matr{X_{Test}}}$ = $\matr{X_{Test}}$ - $\matr{X_{Ref}}$ and predicts the mismatched CSI $\overline{\hat{\matr{H_{Test}}}}$. Finally, the test CSI is predicted with the following addition operation, $ \hat{\matr{H}}_{Test} = \matr{H_{Ref}} + \overline{\hat{\matr{H_{Test}}}}$. Fig.\ref{fig:Proposed Model} illustrates the process of the proposed algorithm. To elaborate, the upper portion in the red dotted box represents the training scheme. Here, the mismatch model $f_{MM}$ learns the appropriate mismatch matrix to modify $\matr{H}_{Ref}$ with respect to  $\matr{X}_{Ref}$. After reaching the optimum point with sufficient iterations, the trained mismatch model shown in the green solid box is used for prediction with the same reference matrices $X_{Ref}$ and $H_{Ref}$. The same reference matrices are used so that the model can use the correlation pattern similar to what it learns in the training phase and can be used in the test phase. The Algorithm \ref{alg:algo1} summarizes the procedures step by step. The offline training defines the training block, and the online testing defines the testing block of Fig. \ref{fig:Proposed Model}. For the offline training scheme, the inputs are the reference matrices $(\matr{X_{Ref}},\matr{H_{Ref}})$, training set ($\matr{X_{Train}}$,$\matr{H_{Train}}$) and $\eta_{threshold}$ to check the convergence. The offline training scheme continues its optimization until $\frac{f_{MM}^{t-1}-f_{MM}^{t}}{f_{MM}^t} > \eta_{threshold}$, where $f_{MM}^{t}$ denotes the average NMSE loss at iteration step $t$. After reaching the optimum point, the well-trained mismatch model will be used in the online estimation.


\subsection{Model Architecture}
\begin{table}[t]
\caption{Architecture of the CNN Model (Single Block)}
    \centering
    \resizebox{0.9\columnwidth}{!}{%
    \begin{tabular}{|c|c|c|c|}
    \hline
      Layer name & Number  & Operations & Filter Size \\
    \hline  
      Input Layer  & 1 & Conv + BN + Relu &  $64 \times (3\times3\times2)$ \\
    \hline
    Middle Layer  & 4 & Conv + BN + Relu &  $64 \times (3\times3\times64)$ \\
    \hline
    Output Layer  & 1 & Conv &  $2 \times (3\times3\times64)$\\
    \hline
    \end{tabular}%
    }
        \label{tab:model_archi}
\end{table}

Our motivation is to verify how the channel correlation can help build the CNN model with a lower computation complexity. 
Following one of the state-of-the-art approaches \cite{liu2021deep}, we propose a CNN model focusing on fewer parameters. Table \ref{tab:model_archi} shows the model parameters of a single block consisting of three layers. More explicitly, we use three blocks that are sequentially connected. Each block refines the previous block's solution. Three different layers are used for each block. All layers have the same filter size, $3 \times 3$. In the input layer, two different channels are used to treat the real part and the imaginary part separately. The middle layer projects the output features from the input layer to the sixty-four channels. At the output layer, it is then again projected into two distinct channels: one is for real numbers, and the other is for imaginary numbers. In each conversational operation, padding is used to keep the dimension of the matrix the same so that the model can be used for any combination of $K$, $M$, and $N$. In the input and middle layers, batch normalization and relu activation, following the convolution operation, are used for better feature selection. Sec.\ref{Sec:Computational Complexity} presents computational complexity comparison with the state-of-the-art model \cite{liu2021deep} and the LS estimator.

\section{Simulation Result}\label{simulation}


To simulate the practical path loss environment and propagation model of an IRS-integrated system, we adapt the simulation setup SimRIS Channel Simulator software \cite{basar2021reconfigurable} to the context of our study. More specifically, we consider an outdoor environment, a $73$ GHz frequency signal, and a cluster of random scattering. As for the fading scheme, the simulation software used the large-scale fading model, assuming that RIS, base stations, and receiver are far from each other. In more detail, it used the free space path loss model as power fall due to the distance and the shadowing model due to the scattered in the environment. All the simulations are done assuming that $K=4$, $M=8$, and $N=32$. The position (x,y,z) of the users, IRS, and the BS are as follows: BS (0,25,20), IRS (70,85,10), User 1 (0,0,0), User 2 (25,10,0), User 3 (40,30,0) and User 4 (20,15,0). The software simulator produces the CSI, $\matr{H_k}$, and Equ.\eqref{target euation} is used to generate noisy received signal $\matr{H_{k}}$. The number of training samples is $60$k, and the number of testing samples (for performance comparison in Sec.\ref{sec:performance_comparison}) is $20$k. For optimizer, Adam optimizer with learning rate $0.0001$ is used. For the loss function, the NMSE (Equ.\eqref{nmse}) is used. Furthermore, we scaled the dataset with a global constant because the channel state values are very low in magnitude, and CNN can not perform well with such a low value. For the reference received signal $\matr{X_{Ref}}$, one additional is generated using the abovementioned parameters. The least-squares (LS) method is used to generate the reference CSI $\matr{H_{Ref}}$ from $\matr{X_{Ref}}$. The impact of choosing different algorithms to generate $\matr{H_{Ref}}$ signal is discussed in Sec. \ref{effect_ref_signal}.

\subsection{Performance Evaluation}
To verify the effectiveness of the proposed model, we use two evaluation metrics, namely, NMSE and spectral efficiency (SE). NMSE captures how the channel estimation diverges from the true value. Here, SE captures the effect of poor channel estimation on designing beam forming vector $\vect{w_k}$ and IRS configuration vector $\vect{r_k}$ \cite{jin2010effect}. We calculate the SE in bps/Hz using the following formula.

\begin{equation} \label{total_SE}
    SE = \frac{\sum_{k=1}^K \log_2(1+\gamma_k)}{K},
\end{equation}
where, $\gamma_k$ is the SNR of user $U_k$. The SNR $\gamma_k$ is given by $\gamma_k = \frac{|\vect{r}_k^H \matr{B}^H_k+\vect{d_k}^H) \vect{w_k}|^2}{ \sum_{i \neq k}^K |(\vect{r}_k^H \matr{B}^H_k+\vect{d_k}^H) \vect{w_i}|^2  + \sigma^2}$,
%
where, $\sigma^2$ is the noise variance of the propagation environment, $\matr{B}_k$ and $\vect{d}_k$ are the perfect channel state information. Beamforming vector $\vect{w}_{k}$ and IRS configuration vector $\vect{r}_{k}$  are optimized, for user $U_k$, based on the our estimated $\hat{\matr{H}}_{Test}$. The beamforming matrix $\matr{W}_{opt}=[\vect{w}_{1},\vect{w}_{2}...\vect{w}_{K}] \in \mathcal{C}^{M \times K}$ is computed by using the alternating optimization method (first Lagrangian dual transform, followed by fractional programming method) described in \cite{guo2019weighted}. Finally,  $\matr{R}_{opt}=[\vect{r}_{1},\vect{r}_{2}...\vect{r}_{K}] \in \mathcal{C}^{N \times K}$ is calculated through a closed form solution by applying the iterative reflection coefficient updating.\\





\subsection{Performance Comparison}\label{sec:performance_comparison}
We determine the NMSE of channel and  SE using \eqref{nmse} and \eqref{total_SE}, respectively varying the strength of the pilot signal from $-5$ dB to $15$ dB. We also determine the SE of the users with exact CSI for reference comparison. All the simulation results are the average of $20000$ training instances. We compare our results with the results of Deep Residual Network (DRN) \cite{liu2021deep} and Least Square Estimate (LS) \cite{biguesh2006training} algorithms.

 \begin{figure}
     \centering
     \includegraphics[scale=0.4]{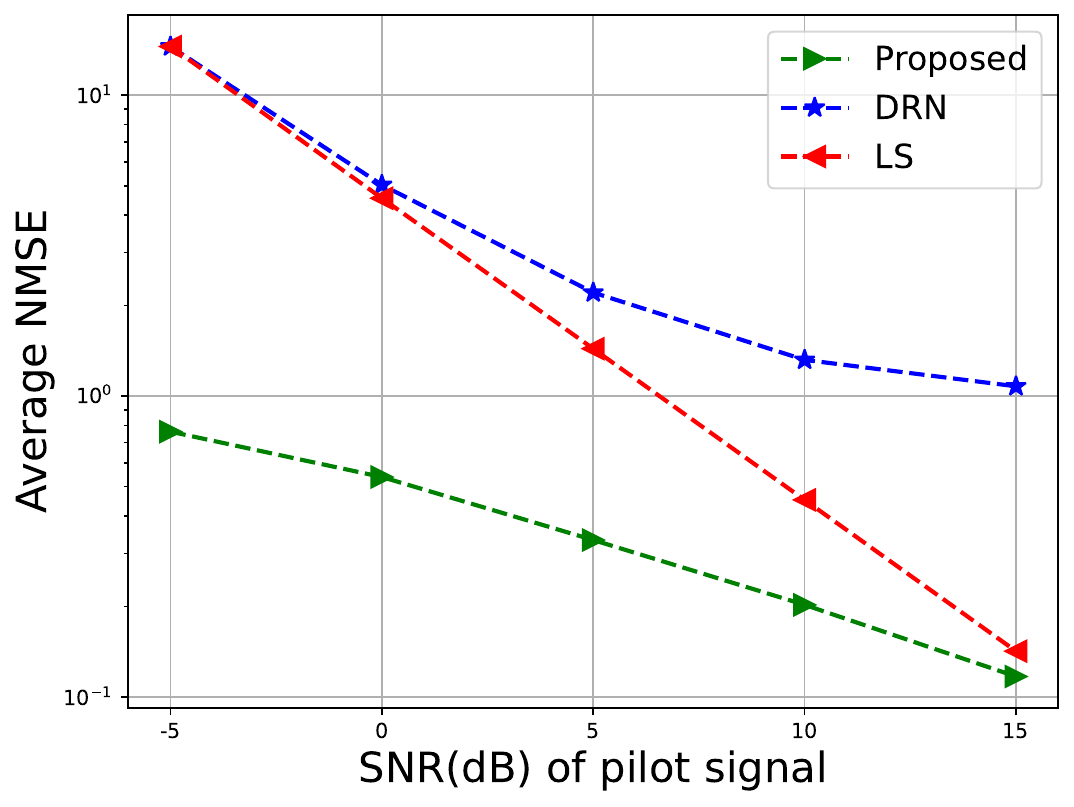}
     \caption{Average NMSE of $4$ distinct users with respect to the pilot signal strength.}
     \label{fig:nms_avg}
 \end{figure}

 \begin{figure}
     \centering
     \includegraphics[scale=0.4]{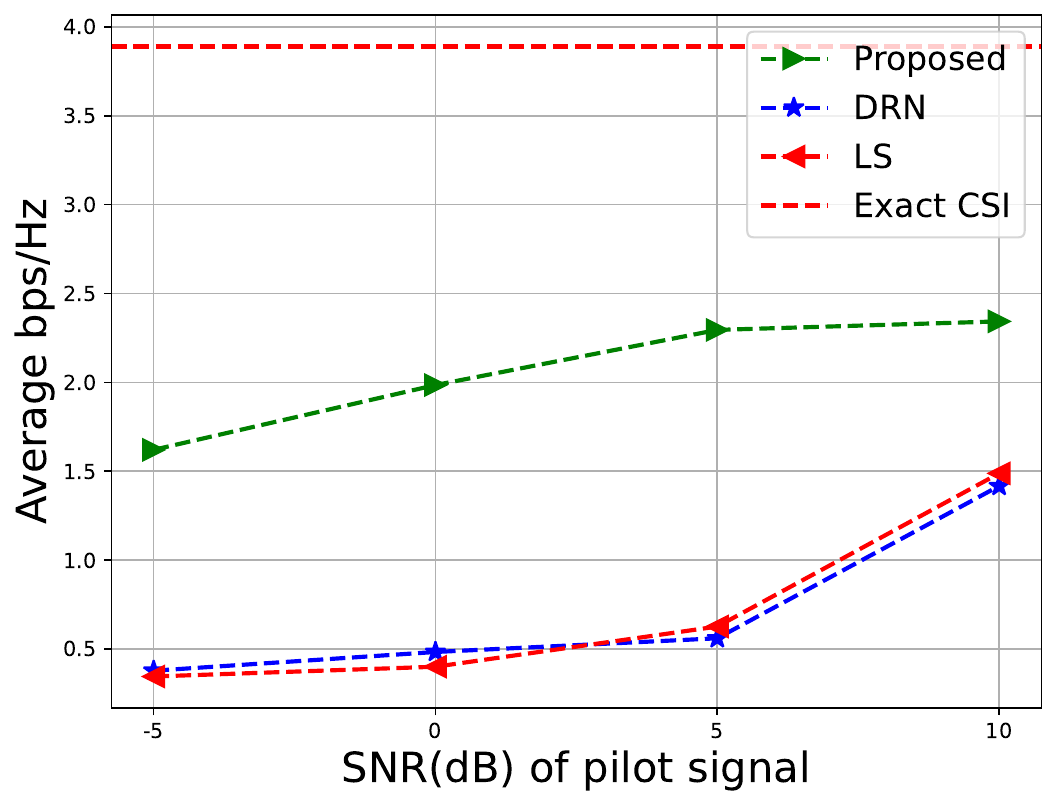}
     \caption{Average spectral efficiency of $4$ distinct users with respect to the pilot signal strength.}
     \label{fig:bps_avg}
 \end{figure}


In Fig. \ref{fig:nms_avg}, we present the NMSE performance for the different algorithms. The results show that at lower strengths of the pilot signal, the NMSE of DRN and LS algorithms are significantly high. In contrast, the NMSE provided by the proposed algorithm is significantly lower. More specifically, the NMSE of both the LS and DRN algorithms are close to $20$ and $7$ for  SNRs of $-5$ dB and $0$ dB, respectively. On the other hand, the NMSE of the proposed algorithm is $0.8$ and $0.6$ for  SNRs of $-5$ dB and $0$ dB, respectively. At the lower SNRs,  NMSE decreases for all the algorithms with increasing the strength of the pilot signal due to the fact that the impact of noise decreases with increasing the strength of the pilot signal. However, the decrement of NMSE for the LS algorithm reduces with increasing the pilot signal strength. At high SNR of pilot strength, both the DRN and proposed algorithm show close NMSE. However, high transmit power is not desirable as it increases interference to the other BSs and users.   
In Fig. \ref{fig:bps_avg}, we present the average SE of the proposed LS and DRN algorithms as well as the SE  with exact CSI. The results clearly show that a significant amount of SE is lost due to the error in channel estimation. We find that the proposed algorithm performs noticeably well compared to the LS and DRN algorithms, especially in the low SNR environment. The spectral efficiency gain is higher than $300\%$ for SNR less than $5$ dB. 

\subsection{Effect of the Reference Signal and Parameters}\label{effect_ref_signal}

   


   


Intuitively, the performance of the proposed model depends on the selection of $\matr{H}_{Ref}$; the higher it represents the actual state of the channel, the better the prediction will be. For instance, NMSE is found to be $0.2665$, $0.34625$, and $0.3325$ with reference signals considered from exact CSI, DRN algorithm, and LS algorithm, respectively, at 5 dB SNR of the pilot signal. As exact CSI is not available in the real-world scenario, an initial good estimation as a reference signal can boost the proposed model's performance.\\
We further analyze the average NMSE using \eqref{nmse}) for different combinations $M$ and $N$ for a fixed user number $K=4$. We use a pilot signal of $5$ dB SNR and exact CSI as the reference signal. For $M=4$ and $,N=16$, the average NMSE is $0.2665$, for $M=16$ and $N=16$, the average NMSE is $0.3217$, and for for $M=16$ and $N=32$, the average NMSE is $0.1143$. It is interesting to observe that the BS antenna size $M$ has an inverse relationship with the performance, where bigger IRS size $N$ boosts the performance.


\subsection{Computational Complexity and Time}\label{Sec:Computational Complexity}

The computational cost for LS is $\mathcal{O}(MC\log_2(MC))$ while for DRN and the proposed algorithm is $\mathcal{O}(D M (N+1) \sum_{l=0}^{N}n_ls_{l+1}^2n_{l+1})$ \cite{liu2021deep}. Here, $D$ is the number of repetition blocks of the middle layer, which is $D=1$ for our proposed method and $D>1$ for DRN, $N_L$ is the number of layers, and $s_l$ is the filter dimension. Thus, the complexity of our proposed method is less than that of the DRN but higher than that of the LS. This can be further evaluated by the computational time to estimate $\overline{\matr{H}_k}$ from \eqref{target euation} for one single CSI. All the simulations were performed using the same machine with specifications: AMD Ryzen $5$ $3600 6$-Core Processor, and $16.0$ GB RAM. We find that the average computation times are 0.011 ms, 0.044 ms, and 0.028 ms for the Ls, DRN, and proposed algorithms, respectively. Although it is expected to lower computation time for the real-case scenario, a high computational resource at the BS can definitely lessen the processing time.

\section{Conclusion and Future Work}
\label{Conclusion}

We have adapted the correlation character among the channel state information into our deep learning model to make a shallow network with minimum parameters. As for the performance matrices, we have considered NMSE and spectral efficiency. In both evaluation metrics, the proposed model shows results comparable to those of the existing approach. Moreover, the proposed algorithm performs better in low SNR regions, which makes it suitable for use in situations with low energy availability or extremely noisy environments. Nevertheless, there is still room to be improved in the high SNR region, which can be a potential research direction. 
\ifCLASSOPTIONcaptionsoff
  \newpage
\fi





\bibliographystyle{IEEEtran}
\bibliography{main}

\vfill


\end{document}